# On the shape of the light profiles of early-type galaxies

N. Caon,[1] M. Capaccioli[2,3] and M. D'Onofrio[2]
[1] *International School for Advanced Studies, Via Beirut 2, 34014 Trieste, Italy*
[2] *Dipartimento di Astronomia, Università di Padova, Vicolo Osservatorio 5, 35122 Padova, Italy*
[3] *Osservatorio Astronomico di Capodimonte, Via Moiariello 16, 80131 Napoli, Italy*

**ABSTRACT**
We have obtained the best fit to the light profiles of a luminosity limited sample of elliptical and S0 galaxies with a power law $r^{1/n}$, letting the exponent remain free rather than keeping it fixed at $1/n = 1/4$ as in the well known de Vaucouleurs formula. The introduction of a free parameter in the fitting formula (ranging from $n = 0.5$ for $\langle r_e \rangle = 0.3$ kpc to $n = 16$ for $\langle r_e \rangle = 25$ kpc) is justified by the existence of a good correlation between $n$ and the global galaxian parameters, such as total luminosity and scale-radius. This result seems to be in line with the segregation of properties between the 'ordinary' and 'bright' families of early-type galaxies, and has consequence for the claimed independence of the shape of galaxy profiles with respect to the Fundamental Plane parameters.

## 1 INTRODUCTION

Several attempts have been made in the past to find 'good' empirical representations of the light profiles of galaxies, to be used as tools for quantitative classification and parametrization, for the identification of components and/or signatures of interactions, and as tests for models and predictions of numerical simulations (cf. the review by Capaccioli 1989).

Among the formulae proposed so far, the $r^{1/4}$ law, introduced by de Vaucouleurs (1948) to represent the light profiles of ellipticals and of the spheroidal components of disc galaxies, has justifiably gained a popularity which has often been confused with 'universality'.

The $r^{1/4}$ formula

$$I(r) = I_e \exp\left\{-k\left[\left(\frac{r}{r_e}\right)^{1/4} - 1\right]\right\} \quad (1)$$

has 'scale' parameters only: a characteristic radius $r_e$, and the corresponding surface brightness $I_e$. The constant $k$ is usually chosen in such a way that $r_e$ is the radius of the isophote containing half of the total luminosity; then $k = 7.6692$, and $r_e$ and $I_e$ are named effective radius and effective surface brightness. Useful formulae and tables related to the $r^{1/4}$ law may be found in Poveda, Iturriaga & Orozoco (1960), de Vaucouleurs (1962), Young (1976), and Mellier & Mathez (1987).

The $r^{1/4}$ formula has succeeded in reproducing, with a remarkable accuracy, the luminosity profiles of quite a few E galaxies. For instance, according to de Vaucouleurs & Capaccioli (1979) and Capaccioli et al. (1990a), the $r^{1/4}$ fit of the surface brightness distribution of the standard E1 NGC 3379 gives observed–calculated (O–C) residuals smaller than 0.08 mag over a 10-mag range. Also, the $r^{1/4}$ law has been often used as a reference for the class of E galaxies in testing and interpreting the results of numerical models of violent relaxation (van Albada 1982), merging of galaxies (Barnes 1988), and tidal stripping of ellipticals (Aguilar & White 1986). While these models produce $r^{1/4}$-like density profiles, a physical basis for this law is still lacking.

Moreover, the analysis of the systematic deviations from the $r^{1/4}$ behaviour has proven an effective tool for the discovery of secondary components (such as stellar discs), substructures (lenses, shells, ripples, etc.) and tidal effects seen as a shallowing or a steepening of the light profiles in the outer parts (cf. Kormendy 1977; Prugniel, Nieto & Simien 1987). Even the very first identification of the two fundamental photometric components of disc galaxies (de Vaucouleurs 1958, 1959) originated from the intuition that bulges of spirals might have $r^{1/4}$ light profiles.

Modern photometry of galaxies has shown, however, that for a large fraction of elliptical galaxies and of spheroids, the $r^{1/4}$ law is only a first-order approximation: often it fits well only a limited interval of the luminosity profile.

For instance, Michard (1985) noted that systematic deviations from the $r^{1/4}$ fit are similar for galaxies of similar luminosities. The best match seems to occur for objects with $M_B = -21$, with typical deviations of the order of 0.1–0.2 mag arcsec$^{-2}$ (Capaccioli 1985).

Schombert (1986) found that, for a large fraction of the brightest cluster galaxies, the $r^{1/4}$ interpolation is applicable to the middle part of the light profile only (between 21 and 25 $B$ mag arcsec$^{-2}$). More importantly, he pointed out that the shape of the light profiles is a unique function of the total luminosity: they become shallower in a $(r^{1/4}, \mu)$ diagram as the luminosity increases (see his fig. 14).

Binggeli & Cameron (1991) have shown that, at the faint end of the galaxy luminosity function, dwarf ellipticals are better described by exponential or by King model profiles. By examining a wider range of luminosities, the same authors conclude that 'profile type is strongly correlated with the total magnitude of the galaxy'.

As for bulges of S0s, it has been noted (e.g. Capaccioli 1987, 1989) that the minor axis light profiles of edge-on objects deviate systematically from the $r^{1/4}$ trend [an upward concavity in the $(r^{1/4}, \mu)$ representation].

All of the above evidence casts some doubts on the universality of the $r^{1/4}$ law, and in particular on its reliability for deriving accurate effective parameters for early-type galaxies, as the results are too dependent on the portion of the light profile that is fitted†. This fact, and the uncertainty of the sky-background subtraction, is the main reason for the large discrepancies in the values of the effective parameters of early-type galaxies reported by different authors (cf. Capaccioli, Caon & D'Onofrio 1992b).

Prompted by these considerations, we decided to study the behaviour of a generalized de Vaucouleurs formula (Sersic 1968), i.e. of a power law in which the exponent is a free parameter $1/n$. Let us say explicitly that the emphasis of this paper is not in showing that the $r^{1/n}$ law provides better fits than the simple de Vaucouleurs formula; improved representations of light profiles are just expected as an obvious consequence of the introduction of a free parameter. Our aim is instead to investigate whether the parameter $n$ has any physical justification by searching for correlations with global photometric parameters (effective radius and total luminosity).

In Section §2 we give a brief account of the properties of the $r^{1/n}$ law. A more complete description (luminosity density, dynamical characteristics and phase-space distribution function) is given by Ciotti (1991). In Section §3 we present the application of the $r^{1/n}$ law to the light profiles of a volume-limited sample of early-type galaxies in the Virgo cluster. Finally, in Section §4 we discuss the implication of the correlation between $n$ and $r_e$, also in relation to the dichotomy between the 'ordinary' and the 'bright' galaxy families claimed by Capaccioli, Caon & D'Onofrio (1993).

## 2  PHOTOMETRIC PROPERTIES OF THE $r^{1/n}$ LAW

In its simplest form the $r^{1/n}$ law, originally proposed by Sersic (1968), is given by the formula

$$I(r) = I_0 \,\text{dex}\bigl(-r^{1/n}\bigr). \tag{2}$$

---

† We remark here that in principle $r_e$ is a model-independent parameter, being simply the radius of the isophote encircling half of the total luminosity. Therefore its identification with the scale-radius of the $r^{1/4}$ law is justified only for those galaxies that are perfectly fitted by the $r^{1/4}$ law over the whole light profile.

Introducing the scale-radius $r_e$, it can be rewritten as

$$I(r) = I_e \,\text{dex}\left\{-b_n\left[\left(\frac{r}{r_e}\right)^{1/n} - 1\right]\right\} \tag{3}$$

and, in logarithmic form, as

$$\mu(r) = \mu_e + c_n\left[\left(\frac{r}{r_e}\right)^{1/n} - 1\right] \tag{4}$$

with $c_n = 2.5\,b_n$. The coefficient $b_n$ can be chosen in such a way that the scale-length $r_e$ is the radius encircling half of the total luminosity $L_T$. The latter can be expressed as a function of $r_e$ and $I_e$

$$L_T = K_n\, I_e\, r_e^2. \tag{5}$$

We have computed the coefficients $b_n$ and $K_n$ by numerical integration of an $r^{1/n}$ galaxy model. In the range explored, $0.5 \leq n \leq 16.5$, they are very well approximated by the relations

$$b_n = 0.868\, n - 0.142 \tag{6}$$

and

$$\log(K_n) = 0.030\,[\log(n)]^2 + 0.441\log(n) + 1.079, \tag{7}$$

both with a rms scatter of 0.014.

For fitting purposes we will use the relation

$$\mu(r) = A + B\, r^{1/n} \tag{8}$$

where, according to equation 4, the effective parameters of the $r^{1/n}$ model are

$$\begin{aligned}\mu_e^\star &= A + c_n, \\ r_e^\star &= (c_n/B)^n.\end{aligned} \tag{9}$$

We have added a superscript $\star$ to the scale parameters of the $r^{1/n}$ formula to prevent confusion with their model-independent analogues.

## 3  AN APPLICATION OF THE $r^{1/n}$ LAW

The $r^{1/n}$ law has been fitted to the major and minor axis light-profiles and to the equivalent profiles of 52 early-type galaxies belonging to a sample of Virgo cluster members that is complete in luminosity: all E and non-barred S0 galaxies of the cluster brighter than $B_T = 14$, or $M_B = -17.3$. The $B$-band photometry that we use here is the result of the coupling of CCD images with deep Schmidt plates. This technique (Capaccioli & Caon 1989) provides accurate light profiles spanning wide surface brightness ranges. More details about the properties of the sample and the data reduction can be found in Caon, Capaccioli & Rampazzo (1990) and in Capaccioli et al. (1992a).

The radial range of the light profiles over which the best fit with relation (8) has been computed goes from outside the region dominated by the seeing (typically a few arcsec, and $\mu_B > 19.6 \pm 1.1$) out to where the uncertainty in the sky subtraction and some possible contamination from nearby objects render the light profiles unreliable. In other words, we fit the formula to the entire range of surface brightness covered by state-of-art photometric observations, removing only what is not measurable or just not directly measured. More specifically, deconvolution experiments (see below and Caon et al. 1990) have been used to determine the boundary of the region where the surface brightness is changed by more than 5 per cent because of seeing convolution. As for the outer boundary of our fitting range, we have adopted the standard value of $\mu_B = 27$ mag arcsec$^{-2}$, where the average photometric error is not larger than 0.25 mag according to fig. 3 in Caon et al. (1990). This limit has been moderately varied to account for the properties intrinsic to each galaxian image.

For some objects with peculiar inner light-profiles, the fit has been restricted to a smaller radial range. In a few cases, especially for very flattened discy galaxies, the fit of the $r^{1/n}$ law was simply impossible. Seven galaxies out of 52 could not be fitted at all because of the presence of prominent disc components, dust lanes or tidal distortions or simply because of the irregular shape of the light profiles (see Table 1). For the remaining 45 galaxies, 12 major axis light profiles and five equivalent profiles could not be fitted by the $r^{1/n}$ law. In some cases the fit required an additional (exponential?) component, while in others the profiles are intrinsically distorted.

In order to minimize the possible repercussions of seeing blurring for results, we have repeated the fitting procedure with light profiles deconvolved according to the technique of Bendinelli (1991), and using in all cases a single Gaussian point spread function (PSF) with the stellar full width at half-maximum (FWHM) measured by Caon et al. (1990).

In practice, for each galaxy profile we have scanned a fine grid of values of $n$, and computed the rms scatter of the (O−C) residuals. We have then chosen as the best value for the exponent the one giving the smallest scatter $\sigma_{bf}$. Note that for many objects the residuals are due to noise and do not reflect systematic departures from the $r^{1/n}$ law. The resulting values of $n$ range from 1 to 15. Two examples of the quality of the fits are shown in fig. 1, for a high and a low value of $n$.

In order to estimate the uncertainty associated with the determination of $n$, for each galaxy (and for the major axis profile only) we computed the values $n_1 < n$ and $n_2 > n$ at which the rms scatter $\sigma$ of the (O−C) residuals exceeds by 25 per cent the value of the best fit. For $n \lesssim 10$ we obtain typically

$$\Delta n = n - n_1 \simeq n_2 - n \simeq 0.25\, n. \qquad (10)$$

For $n > 10$, the above relation still holds for $n_1$, while $n_2 - n$ can reach values as large as $n$ itself. For example, if a light profile is best fitted by an $r^{1/4}$ law (typical scatter $\sigma_{bf} \simeq 0.05$), the use of an $r^{1/3}$ or an $r^{1/5}$ laws would cause $\sigma$ to increase by 25 per cent. On the other hand, when $n > 8$ is the best-fitting result, the same increment of $\sigma$ is achieved by using an $r^{1/(8\pm 2)}$ law.

For each galaxy of our sample, Table 2 provides the identification (column 1), the morphological type according to Binggeli, Sandage & Tammann (1985) (column 2), the total luminosity in the $B$-band, the logarithm of the (model-independent) effective radius $r_e$ (in kpc), and the corresponding surface brightness $I_e$ (columns 3 to 5) from Caon et al. (1990) and Trevisani (1991), the borders $\mu_{start}$ and $\mu_{end}$ of the surface brightness range of the fit (columns 6 and 7), the values of $n$ for the major and minor axes ($n_{maj}$ and $n_{min}$) and for the equivalent light-profile ($n_{eq}$) (column 8), the rms scatter $\sigma_{bf}$ of the fit (column 9), the coefficients

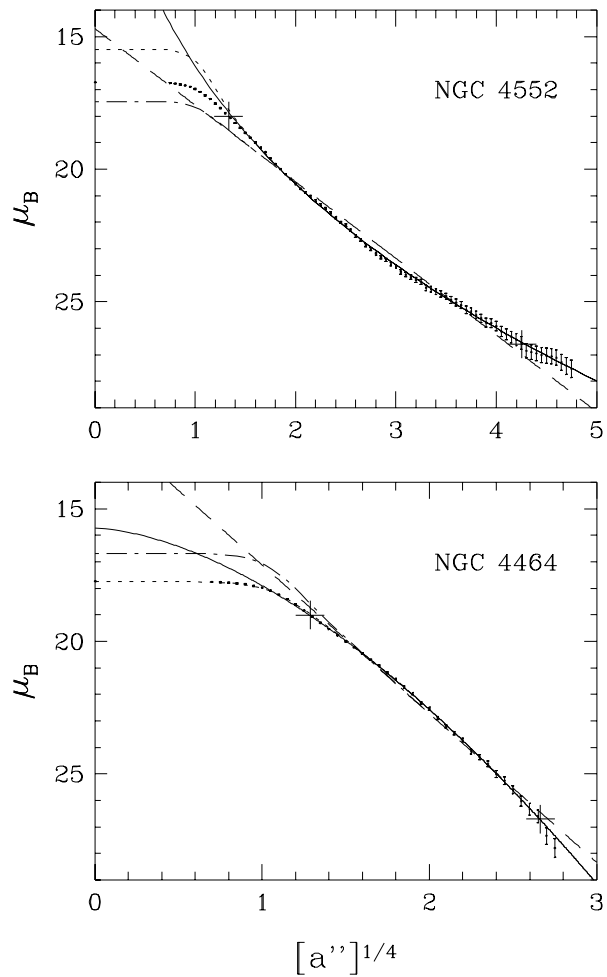

**Figure 1.** Observed major axis $B$-band light profiles of the S0 galaxy NGC 4552 and of the E3 NGC 4464 (filled circles), fitted by an $r^{1/n}$ law (solid line) with $n = 13.9$ and $n = 2.4$ respectively. Crosses mark the borders of the fitting interval. The relative uncertainty of the data is indicated by the error bars, barely visible everywhere but in the outer parts. The dashed lines reproduce the $r^{1/4}$ laws best fitting the data within the same interval. The dotted and dot-dashed lines represent the results of the convolution of the $r^{1/n}$ and the $r^{1/4}$ models with a single Gaussian PSF and the proper FWHM (2 arcsec for both galaxies, according to Caon et al. 1990).